\documentclass[10pt,twocolumn,twoside]{IEEEtran}
\usepackage[pdftex]{graphicx}
\usepackage{cite}
 \usepackage{epstopdf}
\usepackage[cmex10]{amsmath}
\usepackage{graphicx}
\usepackage{slashbox}
\usepackage{mathrsfs}
\usepackage{multirow, multicol, array, comment}
\usepackage[caption=false,font=footnotesize]{subfig}
\usepackage{float}
\usepackage{gensymb,stmaryrd}
\usepackage{algorithmicx}
\usepackage[ruled]{algorithm}
 \usepackage{algpseudocode}
\usepackage{subfig}
  \usepackage{cases}
  \usepackage{hyperref}
 \usepackage{array}  
 \usepackage{verbatim}
 \usepackage{enumerate}
 \usepackage{enumitem}
  \newcolumntype{P}[1]{>{\centering\arraybackslash}p{#1}}
  \newcolumntype{M}[1]{>{\centering\arraybackslash}m{#1}}
\begin{document}
 \title{Epoch-Synchronous Overlap-Add (ESOLA) for Time- and Pitch-Scale Modification of Speech Signals}
 \author{Sunil~Rudresh,~Aditya~Vasisht,~Karthika~Vijayan,~and~Chandra~Sekhar~Seelamantula,~\IEEEmembership{Senior Member,~IEEE} \thanks{S. Rudresh and C. S. Seelamantula are with the Department of Electrical Engineering, Indian Institute of Science (IISc), Bangalore-560012, India (Email: sunilr@iisc.ac.in, chandrasekhar@iisc.ac.in; Phone: +91 80 2293 2695, Fax: +91 80 2360 0444).  A. Vasisht is now with Intel Technology India Pvt Ltd, Bangalore, India (Email: aditya.vasisht@gmail.com). K. Vijayan is now with the Department of Electrical and Computer Engineering, National University of Singapore (NUS), Singapore (Email: karthikavijayan@gmail.com).}}

 \maketitle
 
 \begin{abstract}
 Time- and pitch-scale modifications of speech signals find important applications in speech synthesis, playback systems, voice conversion, learning/hearing aids, etc.. There is a requirement for computationally efficient and real-time implementable algorithms. In this paper, we propose a high quality and computationally efficient time- and pitch-scaling methodology based on the glottal closure instants (GCIs) or epochs in speech signals.  The proposed algorithm, termed as {\it {\textbf {epoch-synchronous overlap-add time/pitch-scaling}}} (ESOLA-TS/PS), segments speech signals into overlapping short-time frames with the overlap between frames being dependent on the time-scaling factor. The adjacent frames are then aligned with respect to the epochs and the frames are overlap-added to synthesize time-scale modified speech. Pitch scaling is achieved by resampling the time-scaled speech by a desired sampling factor. We also propose a concept of epoch embedding into speech signals, which facilitates the identification and time-stamping of samples corresponding to epochs and using them for time/pitch-scaling to multiple scaling factors whenever desired, thereby contributing to faster and efficient implementation. The results of perceptual evaluation tests reported in this paper indicate the superiority of ESOLA over state-of-the-art techniques. The proposed ESOLA significantly outperforms the conventional pitch synchronous overlap-add (PSOLA) techniques in terms of perceptual quality and intelligibility of the modified speech. Unlike the waveform similarity overlap-add (WSOLA)  or synchronous overlap-add (SOLA) techniques, the ESOLA technique has the capability to do exact time-scaling of speech with high quality to any desired modification factor within a range of $0.5$ to $2$. Compared to synchronous overlap-add with fixed synthesis (SOLAFS), the ESOLA is computationally advantageous and at least three times faster. 
 \end{abstract}
 
 \section{Introduction}
 \label{sec:intro}
Time- and pitch-scaling of speech are important problems in speech signal processing. They are relevant in a myriad of applications in speech processing including, but not limited to, speech synthesis, voice conversion, automatic learning aids, hearing aids, voice mail systems, multimedia applications, etc.. The modifications of time duration, pitch, and loudness of speech signals in a controlled manner result in prosody alteration \cite{QuatieriProsody}. Duration expansion and compression are widely used in playback systems, tutorial learning aids, voice mail systems, etc. for slowing down speech for better comprehension or fast scanning of recorded speech data \cite{KSRao}. Altering pitch finds applications in voice conversion systems, animation movie voiceovers, gaming, etc.. Time- and pitch-scale modification are crucial in concatenative speech synthesis, where it is required to manipulate the pitch contours and durations of speech units before concatenating them and later in their post processing. Hence, there is a need for reliable, computationally efficient, and real-time implementable time- and pitch-scale modification techniques in speech signal processing. 

\indent The existing techniques in literature could be broadly classified into two categories, namely, (i) pitch-blind and (ii) pitch synchronous techniques. Next, we briefly present an overview of these two classes of techniques.
 
 \subsection{Pitch-Blind Overlap-Add Techniques}
 \label{sec:pitch_blind_tech}
\subsubsection{Overlap and Add (OLA)} The early techniques relied on simple overlap-add (OLA) algorithms \cite{allen1977unified, allen1977short}, wherein the speech signal is segmented into overlapping frames with an analysis frame-shift of $S_a$. Subsequently, the time-scale modified speech is synthesized by overlap-adding the successive frames after altering the synthesis frame-shift to $S_s=\alpha S_a$, where $\alpha$ is the time-scale modification factor. The analysis frame-shift $S_a$ signifies the number of samples in the frame of speech being processed. The  synthesis frame-shift $S_s$ represents the number of samples of the time-scaled speech synthesized with each overlap-add. The major disadvantage associated with OLA is that it does not guarantee pitch consistency and hence introduces significant artifacts upon time-scale modification. 
 
\subsubsection*{Synchronous Overlap and Add (SOLA)}The synchronous OLA (SOLA) was proposed to introduce some criteria so as to choose which portions of the speech segments must be overlap-added. In SOLA, the successive frames are aligned with each other prior to overlap-add \cite{sola}. The alignment of frames was accomplished using autocorrelation analysis. The speech signal is segmented into overlapping frames with an analysis frame-shift $S_a$ and synthesis frame-shift $S_s=\alpha S_a$, similar to the OLA algorithm. The synthesis frame-shift for each frame is computed such that the successive frames overlap at the locations of maximum waveform similarity between the overlapping frames. That is, the synthesis frame-shift for $i^\text{th}$ frame is altered as $S_s^{(i)}=S_{s}^{(i)}+k_i-k_{i-1}$, where $k_i$ is the offset assuring the frame alignment in a synchronous manner for the $i^{\text{th}}$ synthesis frame and is computed as $k_i=\text{arg}\,\underset{k}{ \text{max}}\,R_i(k)$ and  $R_i(k)$ is correlation between the analysis and synthesis frames under consideration. The drawback of SOLA algorithm is the variable synthesis frame length, i.e., the amount of overlap between successive frames varies for each synthesis frame depending on the correlation between the overlapping frames. The variable length of synthesis frames may not allow for exact time-scaling.  Also, the SOLA algorithm necessitates computation of the correlation function at each synthesis frame, which is computationally expensive.
 
 \subsubsection*{Variants of SOLA} Many variants of the SOLA algorithm have been  proposed to reduce the computational complexity and execution time, mainly by replacing the correlation function with unbiased correlation \cite{379979}, simplified normalized correlation \cite{lawlor}, average maximum difference function (AMDF) \cite{319366},  mean-squared difference function \cite{Wong03fastsola-based},  modified envelope matching \cite{5745327}, etc.. Instead of computing correlation function, a simple peak alignment technique was used to locate the optimum overlap between successive frames of speech having maximum waveform similarity \cite{1198877}. As peak amplitudes can get easily affected by noise, the perceptual quality of the time-scaled speech is highly susceptible to noise. Another variant of SOLA called synchronized and adaptive overlap-add (SAOLA) was proposed, which allows for variable analysis frame length $S_a$ unlike SOLA. SAOLA adaptively chooses $S_a$ as a function of the time-scale modification factor thus reducing the computational load for lower time-scale modification factors \cite{saola}. These algorithms generally perform faster than SOLA, but suffer from reduced quality of time-scaled speech \cite{Dorran06acomparison}.
 
\subsubsection*{SOLA with Fixed Synthesis (SOLAFS)} A significant variant of SOLA termed as SOLA-fixed synthesis (SOLAFS) uses fixed synthesis frame length, instead of variable synthesis frame length, resulting in an improved  quality of time-scale modification. SOLAFS segments the speech signal at an average rate of $S_a$ \cite{solafs}. It allows the beginning of each analysis frame to vary within a narrow interval, such that the adjacent frames of the output speech are aligned with each other in terms of waveform similarity. To be specific, the offset $k_i$ corresponding to maximum waveform similarity affects the beginning point of frames. This flexibility in altering the beginning points of analysis frames facilitates to have a fixed synthesis frame-shift $S_s$, which aids in attaining the exact time-scaling factor. Even though SOLAFS reduces the computational load of SOLA by keeping a fixed synthesis frame rate, it still relies on correlation between two consecutive frames as a measure of waveform similarity.
\subsection{Pitch Synchronous Techniques}
\label{sec:pitch_sync_tech}
Another widely used class of techniques for time- and pitch-scaling is the pitch synchronous overlap-add (PSOLA) \cite{Moulines}, which employs pitch synchronous windowing to segment speech signals. The windowed segments containing at least one pitch period are replicated or discarded appropriately to accomplish required time-scaling. On the other hand, the pitch periods in the windowed segments are resampled by a required factor to achieve pitch-scale modification. The time/pitch-scaled speech signals are synthesized by overlap-adding the modified segments. For PSOLA to provide high quality time/pitch-scaled speech signals, accurate pitch marks, on which the pitch synchronous windows have to be centered, are essential. Inaccurate pitch marks will result in spectral, pitch, and phase mismatches between adjacent frames \cite{Dutoit}. While the time-domain PSOLA (TD-PSOLA) methods operate on the speech waveform itself, frequency-domain PSOLA (FD-PSOLA) methods operate in the spectral domain and are employed only for pitch scaling \cite{MOULINES1995175}. 
 
 \subsubsection*{Linear Prediction PSOLA (LP-PSOLA)}  Application of PSOLA technique on linear prediction (LP) residual \cite{makhoul_LP} results in LP-PSOLA \cite{Moulines, MOULINES1995175}. Accurate pitch markers are required for LP-PSOLA to minimize pitch and phase discontinuities. A recent technique by Rao and Yegnanarayana \cite{KSRao} derives epochs from the LP residual signal of speech and modifies the epoch sequence according to a desired time-scale factor. Then, a modified LP residual is derived from the modified epoch sequence, which is passed through the LP filter to synthesize the time-scaled speech.
 
 \subsubsection*{Waveform Similarity SOLA (WSOLA)} Another technique, which relies on the pitch marks for overlap-add is waveform similarity based SOLA (WSOLA). In WSOLA, the instants of maximum waveform similarity are located using the signal autocorrelation and are used as pitch marks \cite{wsola}. This technique is not capable of producing speech signals with exact time-scale factor due to ambiguities in  replication/deletion of pitch periods chosen based on autocorrelation function. Apart from the autocorrelation, the absolute differences between adjacent frames of speech at different frame-shifts are computed to identify points of maximum waveform similarity \cite{Laprie, veldhuis, mattheyes}.
 
Other considerably different classes of algorithms for time- and pitch-scale modification represent speech in its parametric form using a sinusoidal model \cite{QuatieriProsody}, harmonic plus noise model \cite{Stylianou}, phase vocoder based techniques \cite{phasevocoder}, speech representation and transformation using adaptive interpolation of weighted spectrum (STRAIGHT) model\cite{Kawahara1999187}, etc.. 
 \begin{table*}[t]
\normalsize
\caption{An Objective Comparison of the Proposed Method With State-of-the-Art Methods.  ($N$ is the Number of Samples in a Frame)}
\centering
\begin{tabular}{P{0.14\linewidth}|P{0.20\linewidth}|P{0.17\linewidth}|P{0.17\linewidth}|P{0.13\linewidth}}
\hline
Technique & Criteria for synchronization   & Is {\bf{exact}} time-scaling attained?  & Computational complexity &Output speech quality \\
 \hline \hline
OLA \cite{allen1977unified, allen1977short} & None & Yes  & $\mathcal{O}(1)$&Poor \\ \hline
SOLA \cite{sola}& Cross-correlation & No  & $\mathcal{O}(N^2)$&Moderate\\ \hline
TD-PSOLA   \cite{Moulines}& Alignment of individual pitch periods & No  &  $\mathcal{O}(N\log{N})$ &Moderate\\ \hline
LP-PSOLA \cite{MOULINES1995175} & Alignment of pitch marks from LP residue in frequency domain& No  &  $\mathcal{O}(N^2)$ &Good\\ \hline
SOLAFS \cite{solafs}& Cross-correlation & Yes  &  $\mathcal{O}(N^2)$&High \\ \hline
ESOLA & Epoch alignment in time domain  & Yes  & $\mathcal{O}(N\log{N})$&Very high\\
\hline
\end{tabular}
\label{tab:comparison}
\end{table*}
\section{This Paper}
In this paper, we propose an algorithm to perform time- and pitch-scaling of speech signals exactly to a given factor using epoch-synchronous overlap-add (ESOLA) technique (Section \ref{sec:algo}).  A given speech signal is divided into short-time segments such that each segment contains at least  three or four pitch periods. The analysis frame-shift is adaptively chosen depending upon the time- or pitch-scale modification factor. Pitch-scaling is performed by first time-scaling the speech signal and then appropriately resampling the resulting time-scaled speech. We also propose the concept of epoch embedding into speech signals, which is done by determining epochs and coding the epoch or non-epoch information into the least-significant bit (LSB) of each sample in its $8$/$16$-bit representation (Section \ref{sec:epoch_embed}). Since epoch extraction has to be done only once and the resultant information about epochs is embedded in the speech signal itself, for subsequent time/pitch-scaling, epoch alignment and overlap-add are the only operations required. This minimizes the computational load and reduces the execution time compared with SOLA and its variants, which require computation of correlation between the two frames for each time-scale factor. The proposed technique delivers high quality time-scale modified speech, while being unaffected by pitch, phase, and spectral mismatches.  In Section \ref{sec:compn}, we present a comparative study of the proposed algorithm with the existing state-of-the-art algorithms by indicating  the key differences in terms of perceptual quality of the resulting speech, computational cost, and execution time. Table~\ref{tab:comparison} gives an objective comparison of different time/pitch-scaling techniques with the proposed ESOLA technique. Since, all the techniques in Table~\ref{tab:comparison} employ frame-based analysis and synthesis, $N$ used in the computational complexity column denotes the number of speech samples in a frame. Section \ref{sec:eval} presents a detailed perceptual evaluation of performances of different time/pitch-scaling algorithms, {\it vis-\`a-vis} the proposed ESOLA technique. In Section \label{sec:implementation}, we discuss about the variation of the ESOLA technique for continuously changing the time/pitch-scale factor for a speech signal. The proposed method has been implemented on various platforms such as MATLAB, Python, Praat, and Android. Time- and pitch-scaled speech signals for a few Indian and Foreign languages, vocals, synthesized speech, speech downloaded from YouTube are put up on the internet for the benefit of readers and can be accessed by the link \href{http://spectrumee.wixsite.com/spectrumtts}{http://spectrumee.wixsite.com/spectrumtts}.
 
\section{Epoch Extraction and its Role in Time- and Pitch-Scaling of Speech Signals}
\subsection{Role of Epochs in Time/Pitch-Scaling}
Voiced speech is produced by exciting the time-varying vocal-tract system primarily by a sequence of glottal pulses. The excitation to the vocal tract system is constituted by the air flow from lungs, which is modulated into quasi-periodic puffs by the vocal folds at glottis. The vibrations of vocal folds (closing and opening the wind pipe at the glottis) acoustically couple the supra-laryngeal vocal tract and trachea. Although glottal pulses are the source of excitation, the significant excitation of the vocal-tract system occurs at the instant of glottal closure. Such impulse-like excitations during the closing phase of a glottal cycle are termed as epochs or glottal closure instants (GCIs) \cite{Rabiner}. The speech thus produced is a quasi-periodic signal with pitch periods characterized by epochs. Pitch is a prominent speaker-specific property and it does not vary largely with the rate of speaking.  An analysis of change in distribution of fundamental frequency ($F_0$) with the change in speaking rate suggests that variation in $F_0$ is speaker specific \cite{epoch_yegna}. That is, some speakers are able to maintain the same $F_0$ at different speaking rates. In other words, they can produce speech at different speaking rates while maintaining intelligibility and naturalness, which is exactly what we seek in time-scale modification of speech. Since, $F_0$ inherently depends on epochs, the very less variation of $F_0$ is attributed to less variation in the pitch periods. This motivates us to use epochs as anchor points for synchronizing consecutive frames for time-scale modification.\\

\subsection{Epoch Extraction Algorithms}
\label{sec:zff}
Determining epochs from speech signals is a non-trivial task and several algorithms have been proposed to solve the problem. Initial attempts were aimed at points of maximum short-time energy in segments of speech \cite{Laprie2, Lian, ewender}. The estimates of the pitch marks obtained using these techniques were refined using dynamic programming strategies, minimizing cost functions formulated based on waveform similarity and sustainment of continuous pitch contours over successive frames of speech \cite{Laprie, Lian, mattheyes, Tsiakoulis}. The drawback of most of these algorithms is the utilization of several adhoc parameters. Epochs have also been obtained by identifying points of maximum energy in Hilbert envelope \cite{389229}, by using group delay function \cite{KSRao,KSRao_GD}, using residual excitation and a mean-based signal
(SEDREAMS) technique \cite{sedreams}, based on spectral zero crossings \cite{szcr}, from positive zero crossings of zero frequency filtered (ZFF) signal \cite{ZF}, based on dynamic plosion index  (DPI) \cite{DPI}, etc.. An extensive review of the various epoch extraction algorithms and their empirical computational complexity has been given in \cite{drugman_review}. Any of these algorithms could be used as long as they give reliable estimates of epochs and are computationally efficient. As reviewed in \cite{drugman_review} and \cite{Adiga_spcom_epoch}, SEDREAMS, ZFF, or DPI give the most accurate estimate of epochs and a version of SEDREAMS called \textit{fast} SEDREAMS is computationally efficient than the rest of the techniques \cite{drugman_review}.

In this paper, we use the zero frequency resonator (ZFR) proposed by Murty and Yegnanarayana \cite{ZF} for epoch extraction as it gives reliable estimates and requires less computational resources for implementation. 
The ZFR filters speech signals at a very narrow frequency band around 0 Hz, as this low frequency band of speech is not affected by the 
vocal tract system. The resulting signal is termed as zero frequency signal and it exhibits discontinuities at epoch locations as positive zero 
crossings \cite{ZF}. The procedure of obtaining epochs using ZFR is summarized below.
\begin{itemize}
\item Speech signal $s[n]$ is preprocessed to remove the low-frequency bias present as $x[n]=s[n]-s[n-1]$.
\item The signal $x[n]$ is passed through an ideal zero-frequency resonator (integrator) two times. This is done to reduce the effect of vocal tract on the resulting signal. 
 $$ y_1[n]=-\sum_{k=1}^2 a_k y_1[n-k]+x[n],$$ 
$$ y_2[n]=-\sum_{k=1}^2 a_k y_2[n-k]+y_1[n].$$
\item The trend in $y_2[n]$ is removed by successively applying a mean-subtraction operation.
$$y[n]=y_2[n]-\frac{1}{2N+1}\sum_{m=-N}^N y_2[n+m]$$
 The value of $2N+1$ is chosen as to lie between $1$ to $2$ times the average pitch period of the speaker under consideration. 
 \item The positive zero crossings of $y[n]$ indicate the epochs.
\end{itemize}

\indent Next, we propose a new time- and pitch-scale modification technique (ESOLA) based on epoch alignment.
 
 \section{ESOLA: Epoch Synchronous Overlap-Add Time- and Pitch-Scale Modification of Speech Signals}
 \label{sec:algo}
Time-scale modification is generally performed by discarding or repeating short-time segments of speech, or by manipulating the amount of overlap between
successive segments. Pitch-scale modification involves resampling of the speech signal. In this paper, we adopt a pitch-blind windowing for segmentation of speech signals into overlapping frames (typically, with $50$\% overlap). Subsequently, the overlap between successive frames are increased or decreased for duration compression or expansion, respectively. Depending on the desired time-scale modification factor, the overlap between successive frames, or equivalently the frame-shift, is modified. The newly formed frames (analysis frames) with modified frame-shifts are overlap-added to 
 synthesize duration-modified speech signals. To perform pitch-scaling, first, the speech signal is time-scaled by an appropriate factor and then resampled to match the length of original speech signal. 
A crucial requirement for time-scaling techniques is pitch consistency, i.e., the pitch of time-scaled speech signals should not vary with duration expansion or compression. We employ epoch alignment as a measure of synchronization between successive speech frames prior to overlap-add synthesis to ensure pitch consistency.

For time-scale modification, the speech signal $x[n]$ is segmented into frames, $x_m[n]$ of length $N$. Generally, $N$ is chosen such that each  frame contains three or four pitch periods. The average frame-shift between successive frames is $S_a$. The exact analysis frame-shift is decided based  on the desired time-scale modification factor. The analysis frames are selected in such a way that the overlap between the successive analysis frames is more when duration of speech  signal has to be increased (slower speaking rate) compared with the amount of overlap when the duration has to be decreased (faster speaking rate). The $m^{\text{th}}$ analysis frame of a speech signal $x[n]$ is given by, 
\begin{eqnarray}
 x_m[n]=x[n+mS_a+k_m], \quad n\in\llbracket 0, N-1 \rrbracket,
 \label{eq:analysis_frame}
\end{eqnarray}
where $\llbracket 0, N-1 \rrbracket$ denotes the integer set $\{0, 1, \cdots, N-1\}$ and $k_m$ is the additional frame-shift ensuring frame alignment for the $m^{\text{th}}$ analysis frame. The synthesis frame-shift is chosen as $S_s=\alpha S_a$, where $\alpha $ is the desired time-scale modification factor. The $m^\text{th}$ synthesis frame is denoted by 
\begin{eqnarray}
 y_m[n]=y[n+mS_s], \quad n\in\llbracket 0, N-1 \rrbracket,
 \label{eq:synthesis_frame}
\end{eqnarray}
where $y$ is the time-scaled signal. Note that the length of both analysis and synthesis frames is $N$ and is fixed. Next, we discuss the frame alignment process, which in turn involves epoch alignment between the frames, which determines $k_m$.
\begin{figure}
  \centering
  \includegraphics[width= \columnwidth]{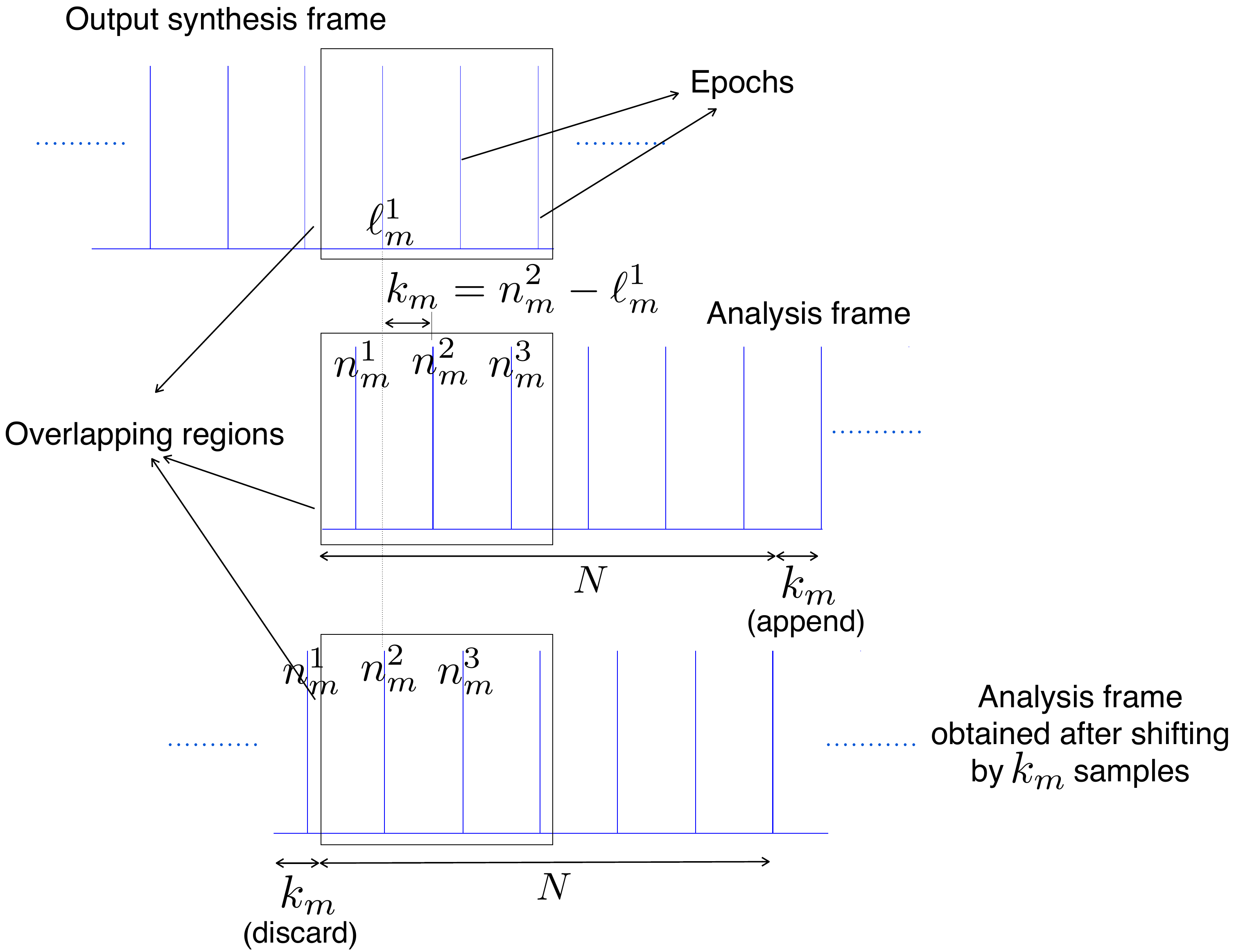}
  \caption{Illustration of epoch alignment between synthesis and analysis frames.}
  \label{fig:epochAlign}
 \end{figure}

\subsection{Epoch Alignment}
\label{sec:epoch_alignment}
\begin{figure}[t]
  \centering
  \includegraphics[width= \columnwidth]{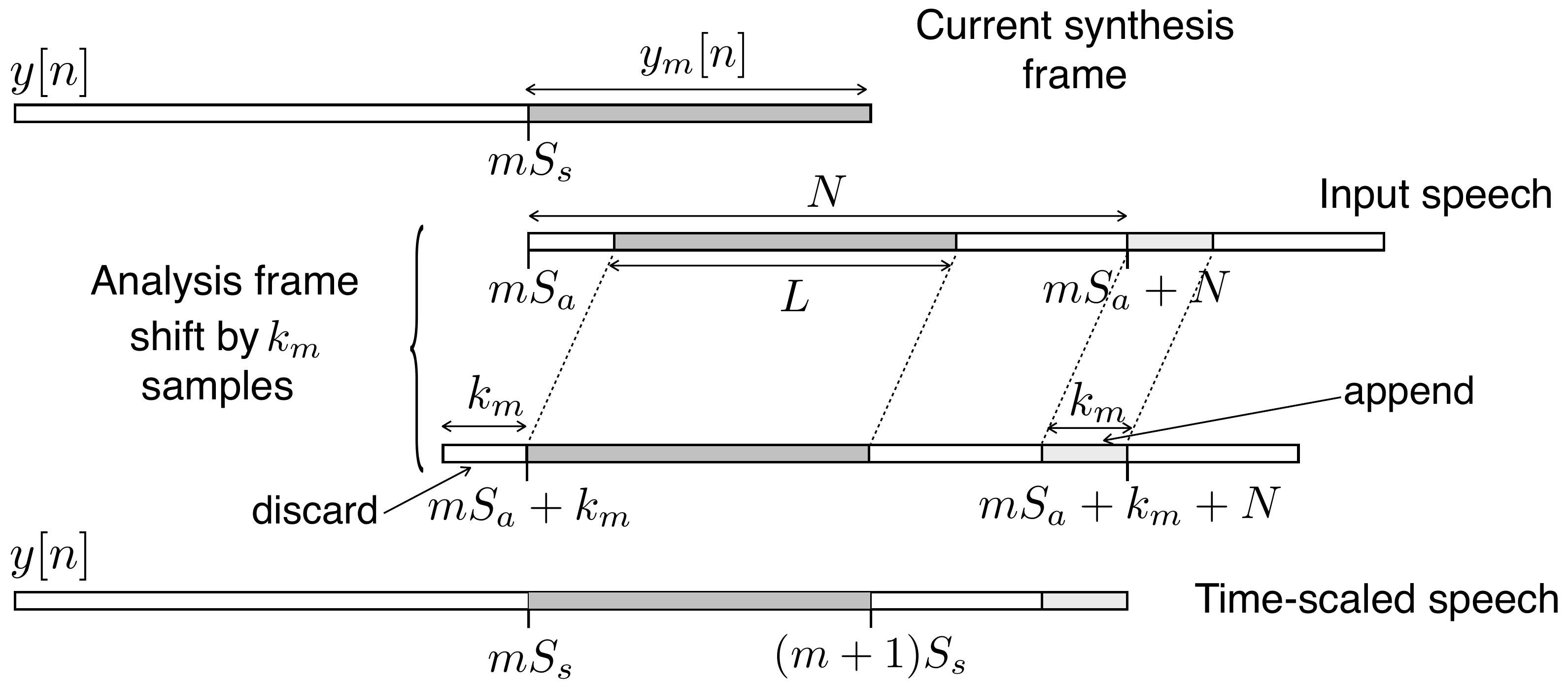}
  \caption{A schematic illustration of ESOLA-TS technique.}
  \label{fig:ETS_schematic_illustration}
 \end{figure}

The process of aligning frames with respect to epochs in order to compute $k_m$ is illustrated in Fig.~\ref{fig:epochAlign}. The $m^{\text{th}}$  analysis frame of speech $x_m[n]$, which begins at $mS_a$ is to be shifted and aligned with the $m^{\text{th}}$ synthesis frame, $y_{m}[n]=y[n+mS_s]$, $n\in\llbracket 0, N-1 \rrbracket$, which begins at $mS_s$ and overlap-added to get the $m^\text{th}$ output frame. Let $\ell_{m}^{1}$ denote the location index of the first epoch in the $m^\text{th}$ synthesis frame $y_m$. Let $\{n_{m}^{1}, n_{m}^{2}, \cdots, n_{m}^{P}\}$ be the indices of $P$ epochs in the $m^\text{th}$ analysis frame $x_m$. Now, the analysis frame-shift $k_m$ is computed as $$k_m=\displaystyle\underset{{1 \leq i \leq P}  } \min \,(n_{m}^{i}-\ell_{m}^{1}),$$ such that $k_m \geq 0$. The shift factor $k_m$ ensures frame alignment by forcing the $m^\text{th}$ analysis frame to begin at $mS_a+k_m$ according to \eqref{eq:analysis_frame} as shown in Fig \ref{fig:ETS_schematic_illustration}. Hence, the first epoch occurring in $y_{m}$ after the instant $mS_s$ is aligned with the next nearest occurring epoch in $x_m$.  Thus, the epochs in the synthesis and modified analysis frames are aligned with each other and any undesirable effects perceived as a result of pitch inconsistencies are mitigated from the time-scale modified speech. 
 \subsection{Overlap and Add}
\label{sec:overlap_add}
\begin{figure}
  \centering
  \includegraphics[width= \columnwidth]{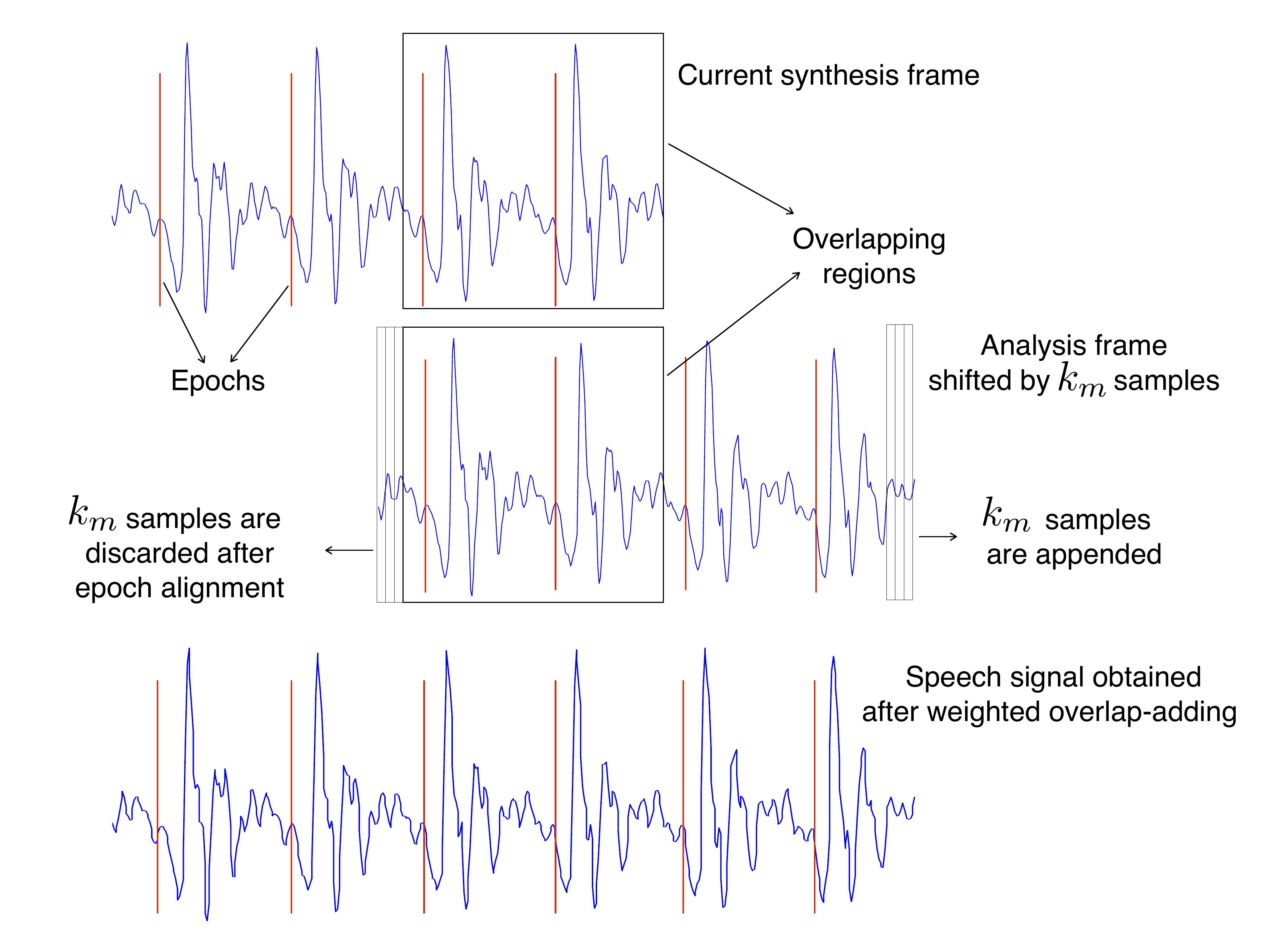}
  \caption{[Color online] Illustration of ESOLA-TS technique on a segment of a speech waveform.}
  \label{fig:wavefromAlign}
 \end{figure}

In order to nullify the possible artifacts due to variable length of overlap-add region, we keep the fixed synthesis length. The additional $k_m$ samples for the $m^\text{th}$ analysis frame, which now begins at $mS_a+k_m$ due to the shift $k_m$ and ends at $mS_a+k_m+N$ are appended from the successive analysis frame, prior to overlap-add synthesis. This is done to ensure that $y_m$ holds exactly the required number of samples as demanded by the time-scale modification factor thereby delivering exact time-scale modification. The analysis shift $k_m$ and overlap-add are done directly on the time-domain speech signal as shown in Fig. \ref{fig:wavefromAlign}. 

\begin{figure}[t]
  \centering
  \begin{tabular}{cc}
   \hspace{-0.2 in} \raisebox{-0.5\height}{ \includegraphics[width=3.3 in]{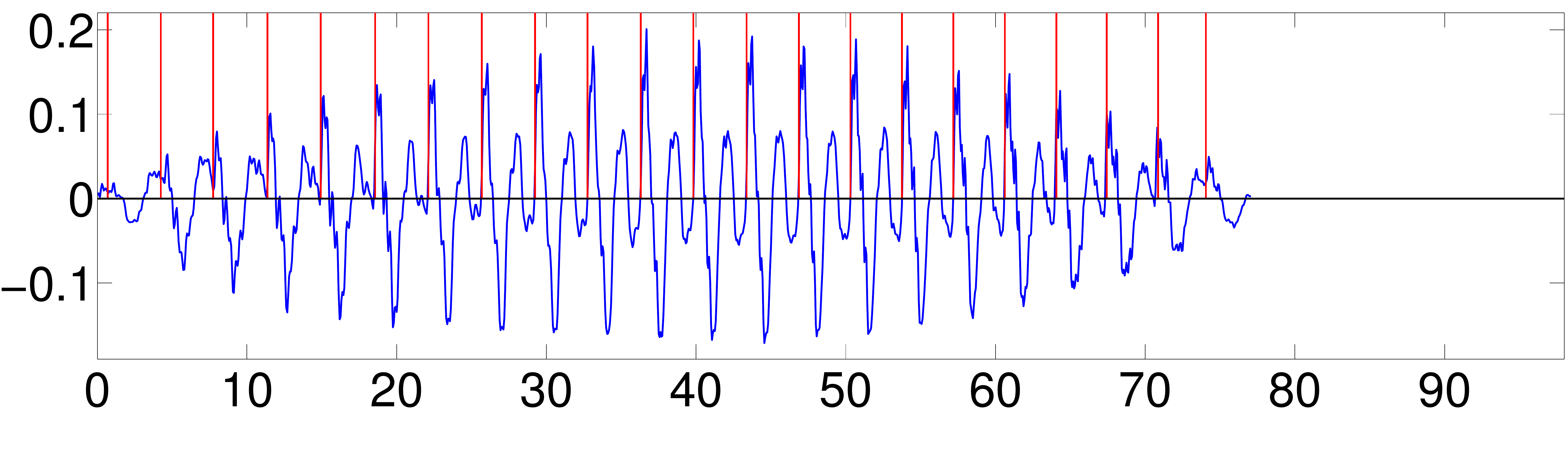}}& \hspace{-0.15 in}{(a)}\\ 
  \hspace{-0.2 in} \raisebox{-0.5\height}{  \includegraphics[width=3.3 in]{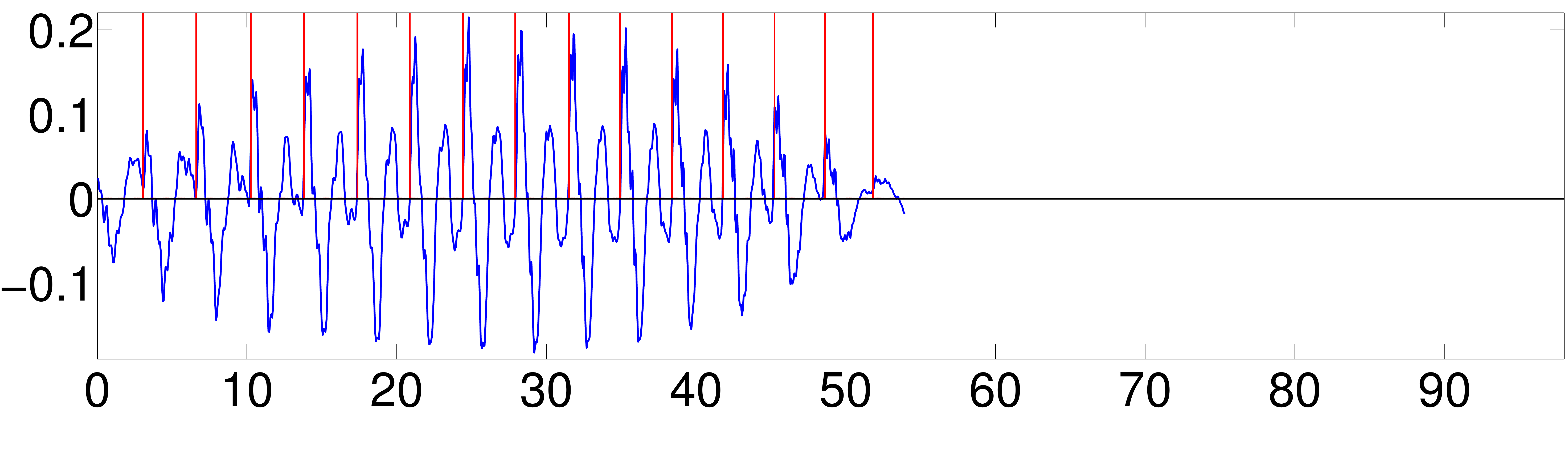}}& \hspace{-0.15 in}(b)\\
 \hspace{-0.2 in}  \raisebox{-0.5\height}{  \includegraphics[width=3.3 in]{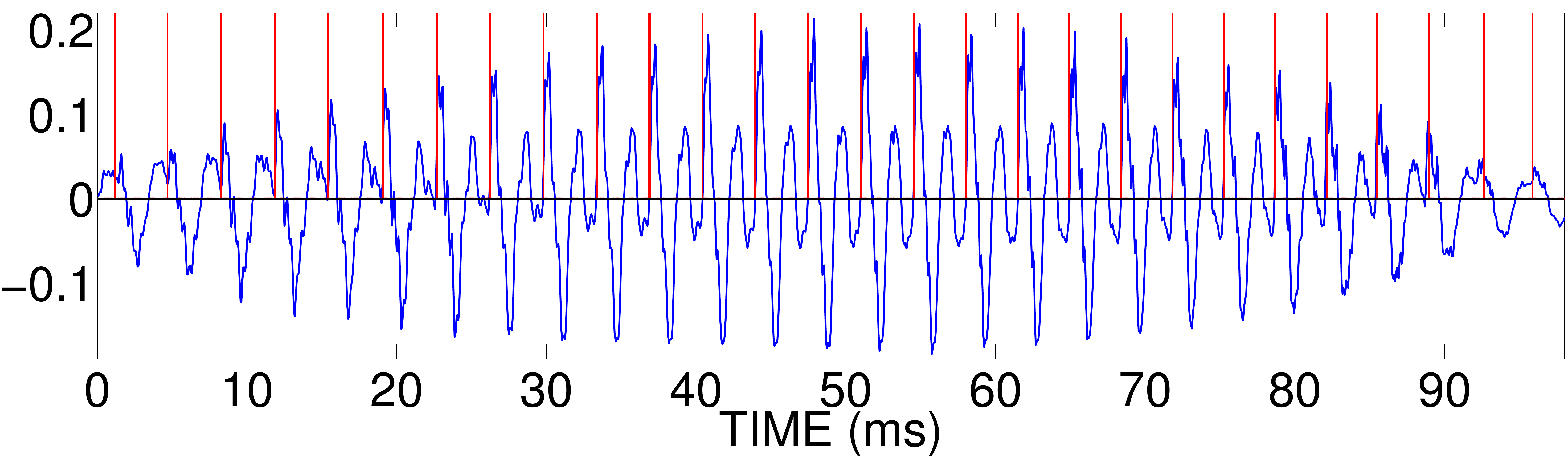}}&\hspace{-0.15 in}(c) \\
  \end{tabular}
    \caption{[Color online] Time-scale modification using the ESOLA technique. (a) Original speech signal; time-scaled signals (b) $\alpha=0.7$ and (c) $\alpha=1.3$. It is observed that the average pitch period in all the three speech segments remains more or less the same.}
  \label{fig:TSM}
 \end{figure}

The modified analysis frame is overlap-added to the current output frame $y_m$ using a set of cross-fading functions $\beta[n]$ and $(1-\beta[n])$ as  given in \eqref{eq:synth_eq}. The fading function $\beta[n]$ could be a linear function or a raised-cosine function employed to reduce audible artifacts due to overlap of two frames during synthesis. The time-scaled output signal is synthesized as
\begin{numcases} { y[n+mS_s] =}
\beta[n]y_m[n] \nonumber \\ 
\,\,+\left(1-\beta[n]\right)x_m[n],\quad 0 \leq n \leq L-1, \nonumber \\
x_m[n], \quad  L \leq n < N,
 \label{eq:synth_eq}
\end{numcases}
where $L= N-S_s$ denotes the overlapping region between the frames. Fig.~\ref{fig:TSM} shows a 
segment of speech and its time-scaled versions to two different scale-factors. It is observed that the proposed ESOLA technique provides high 
quality time-scaled speech signals with pitch consistency, thereby preserving speaker characteristics.
\subsection{Selection of Parameters}
\label{sec:parameters}
\subsubsection*{Range of $k_m$} 
In voiced regions, both analysis and synthesis frames contain valid epochs and $k_m$ is computed as described in Section \ref{sec:epoch_alignment}. In the case of unvoiced regions, epoch extraction algorithms may give spurious epochs and $k_m$ is computed in the same way as that for voiced regions. Since, there are no pitch periods present in unvoiced regions, the epoch alignment doesn't make sense. In other words, carrying out the process of aligning the frames using spurious epochs in unvoiced regions doesn't create any pitch inconsistency in the time-scaled signal. However, there might be cases where no epochs are present in one of the two frames or both the frames. In these cases, analysis shift $k_m$ is set to zero, i.e., the frames are overlap-added without any alignment. In the extreme case, the maximum value of $k_m$ is set to $k_{max}$, which is equal to the synthesis frame-shift $S_s$. Thus the range of analysis shift is given by $0 \leq k_m \leq k_{max} (=S_s)$.  
\subsubsection*{Selection of $N$, $S_a$, and $S_s$}
The length of analysis and synthesis frames is fixed as $N$. Typically, the frame length $N$ is chosen to contain at least  three or four pitch periods/epochs. In this paper, we have used $20$ ms frame length, which gives $N=20\times 10^{-3} F_s$, where $F_s$ is the sampling frequency. Since the length of synthesis frame is set to $N$, the synthesis rate $S_s$ depends on the amount of overlap ($L$) and is given by $S_s = N-L$. In our experiments, the amount of overlap $L$ is chosen to be $50$\%, which gives $L=N/2$ and hence $S_s=N/2$. Also, the analysis frame rate is related to synthesis frame rate as $S_s=\alpha S_a$.
  \begin{figure}
  \centering
  \begin{tabular}{cc}
   \hspace{-0.2 in} \raisebox{-0.5\height}{ \includegraphics[width=3.3 in]{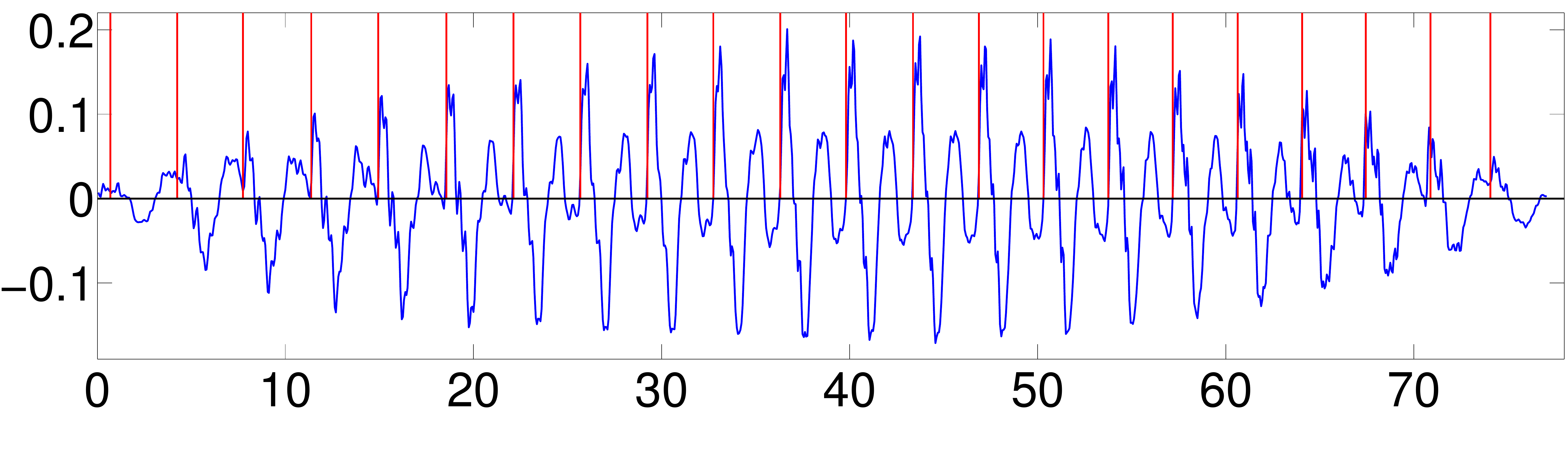}}& \hspace{-0.15 in}{(a)}\\ 
  \hspace{-0.2 in} \raisebox{-0.5\height}{  \includegraphics[width=3.3 in]{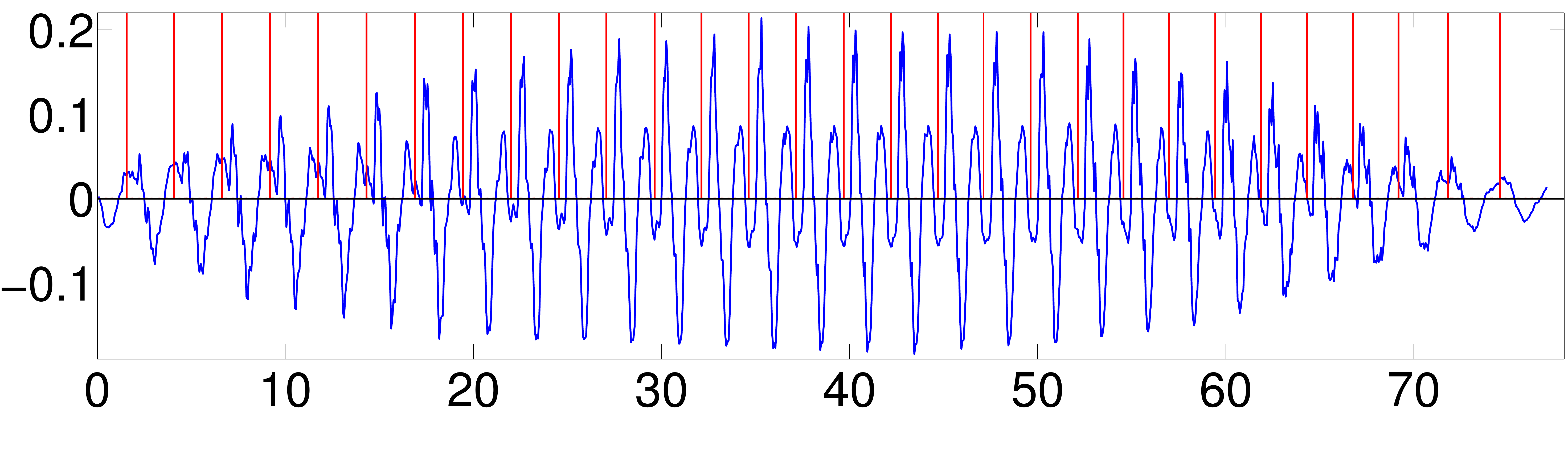}}& \hspace{-0.15 in}(b)\\
 \hspace{-0.2 in}  \raisebox{-0.5\height}{  \includegraphics[width=3.3 in]{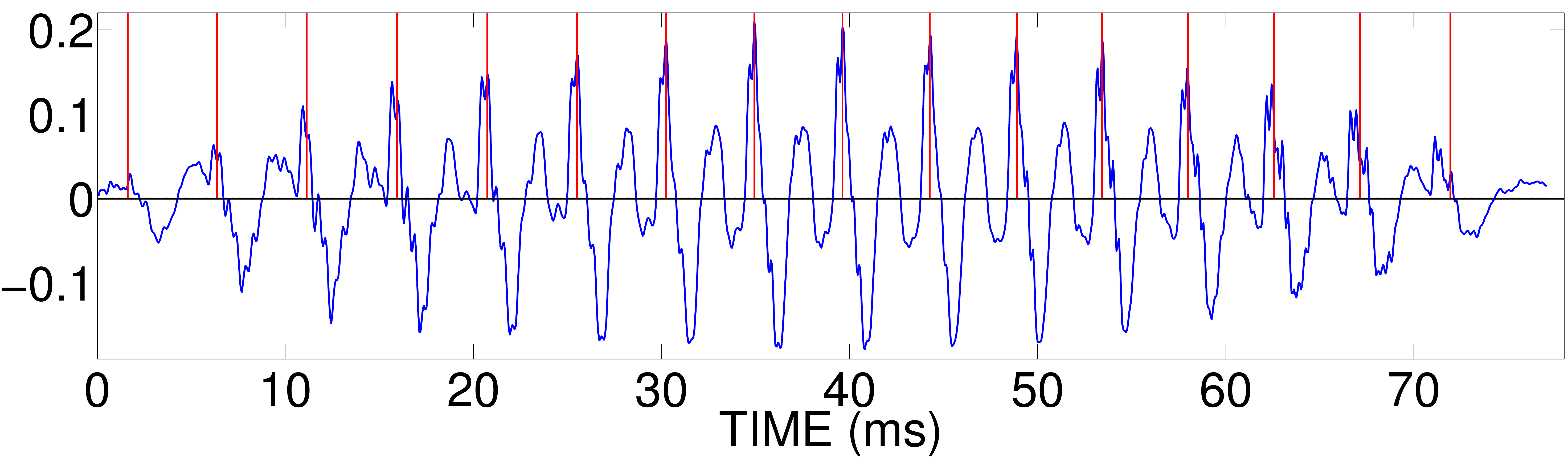}}&\hspace{-0.15 in}(c) \\
  \end{tabular}
  \caption{[Color online] Pitch-scale modification using the ESOLA technique. (a) Original speech signal; pitch-scaled signals (b) $\beta=1.4$ and (c) $\beta=0.75$. It is observed that the average pitch period changes according to the scaling factors but the duration of all the three segments remains the same.}
  \label{fig:PSM}
 \end{figure}
\subsection{Pitch-Scale Modification}
Resampling of a speech signal alters both pitch and duration of the signal. As we have an efficient time-scaling technique, it can be employed for pitch-scaling.  For a given pitch modification factor ($\beta$), first, the speech signal is time-scaled by a factor, which is the reciprocal of the pitch-scale factor ($1/\beta$) and then the time-scaled speech is appropriately resampled (by the factor $F_s/\beta$) to match the length of the resampled signal to that of the original speech signal. Thus, the pitch-scaled signal has a different pitch because of the resampling and the length is unaltered due to time-scaling. It has been observed that the output speech stay consistent even if the order of these two operations gets reversed, i.e., the pitch-scale modification is invariant over the order in which time-scaling and resampling  are performed. We observe that the resulting pitch-scaled signal is of high quality devoid of any artifacts as compared with the other techniques such as PSOLA. Fig.~\ref{fig:PSM} shows the pitch-scale modification of a segment of speech signal to two different modification factors.

\begin{figure}
  \centering
  \begin{tabular}{cc}
   \hspace{-0.2 in} \raisebox{-0.5\height}{ \includegraphics[width=3.3 in]{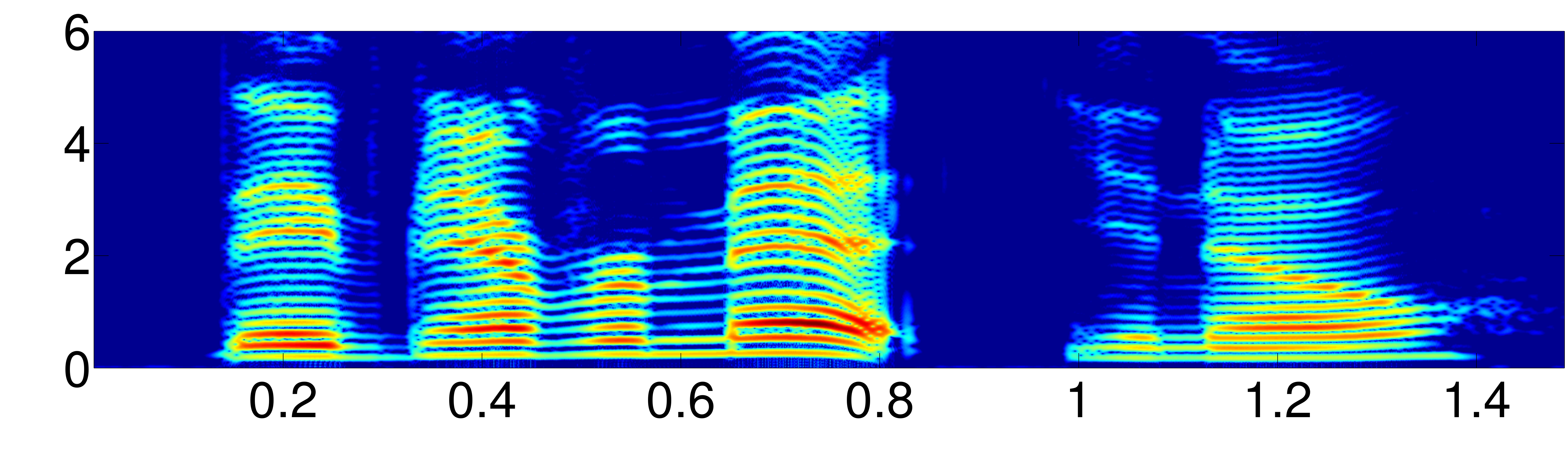}}& \hspace{-0.15 in}{(a)}\\
  \hspace{-0.2 in} \raisebox{-0.5\height}{  \includegraphics[width=3.3 in]{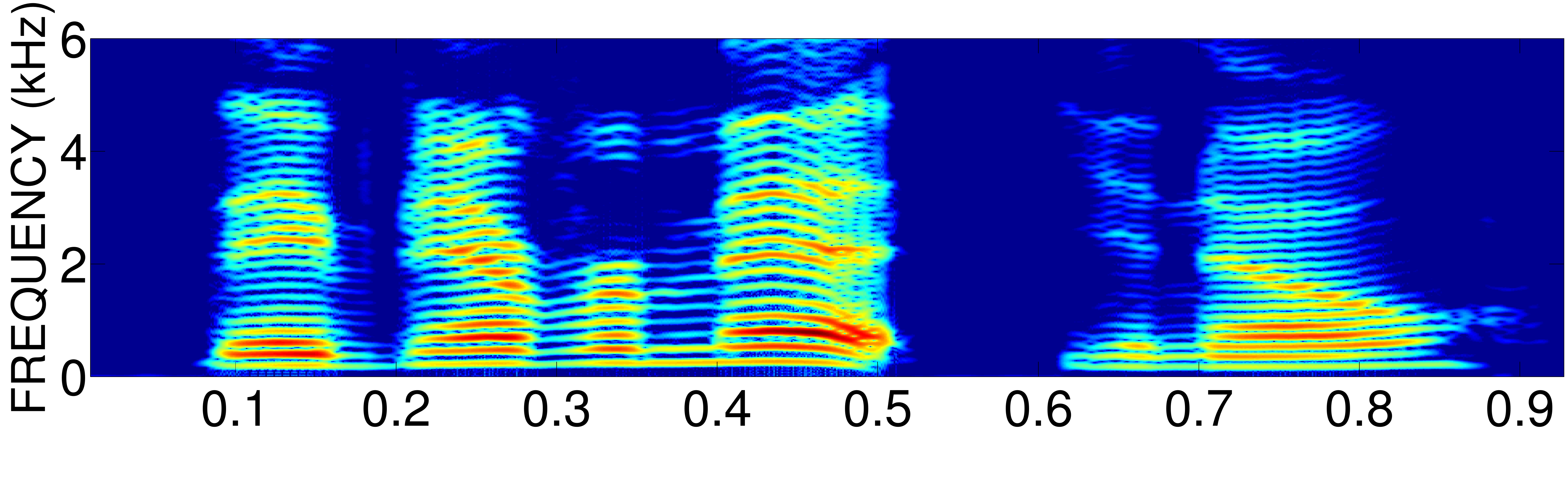}}& \hspace{-0.15 in}(b)\\
 \hspace{-0.2 in}  \raisebox{-0.5\height}{  \includegraphics[width=3.3 in]{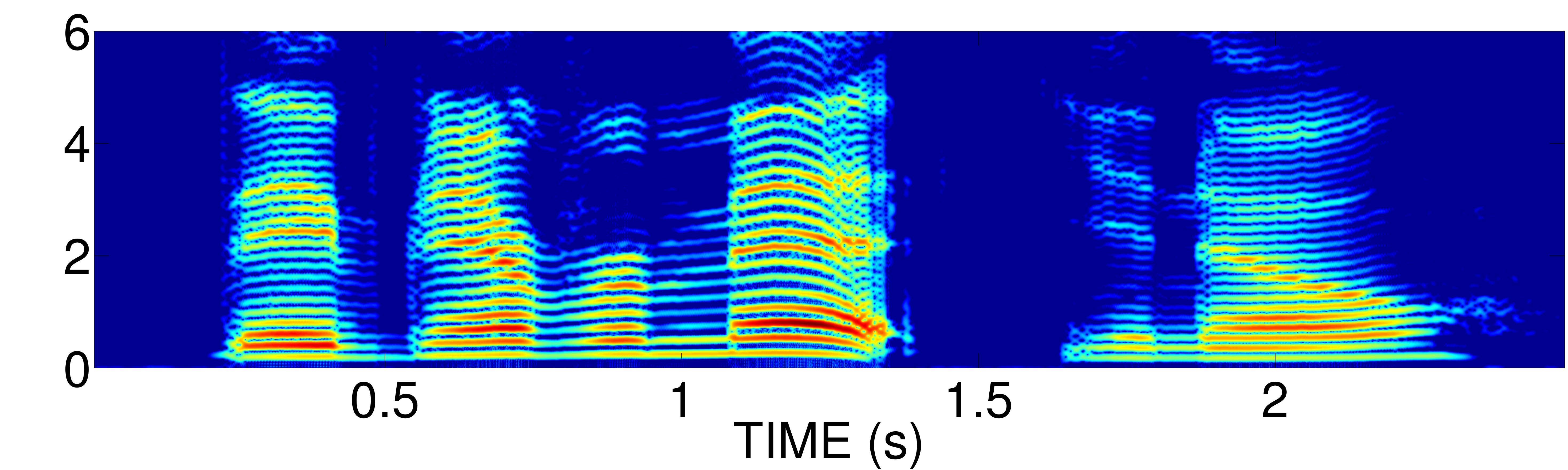}}&\hspace{-0.15 in}(c) \\
  \end{tabular}
  \caption{[Color online] Spectrograms of the utterance ``they never met, you know''. (a) Original speech signal; time-scaled signals for (b) $\alpha= 0.625$ and (c) $\alpha=1.66$.}
  \label{fig:TSM_result_spec}
 \end{figure}

\begin{figure}
  \centering
    \begin{tabular}{ll}
  \hspace{-0.2 in} \raisebox{-0.5\height}{ \includegraphics[width=3.3 in]{spectrogram_ETS_1.pdf}} \hspace{-0.03 in}{(a)}\\
 \hspace{-0.2 in} \raisebox{-0.5\height}{  \includegraphics[width=3.3 in]{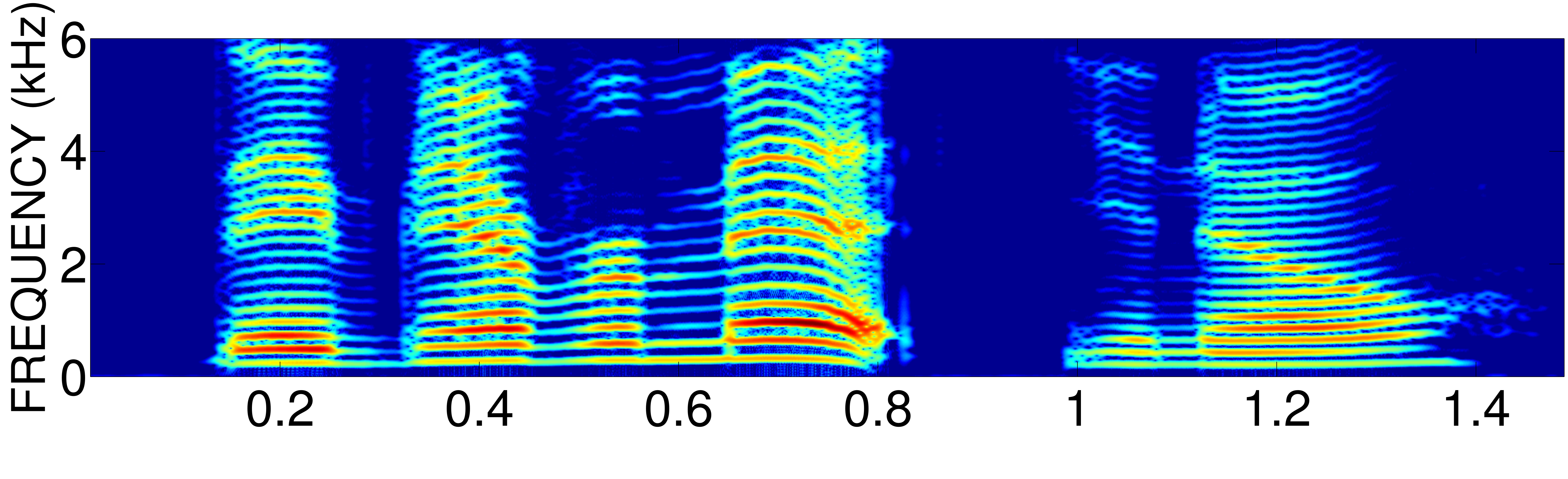}} \hspace{-0.03 in}{(b)}\\
 \hspace{-0.2 in} \raisebox{-0.5\height}{  \includegraphics[width=3.3 in]{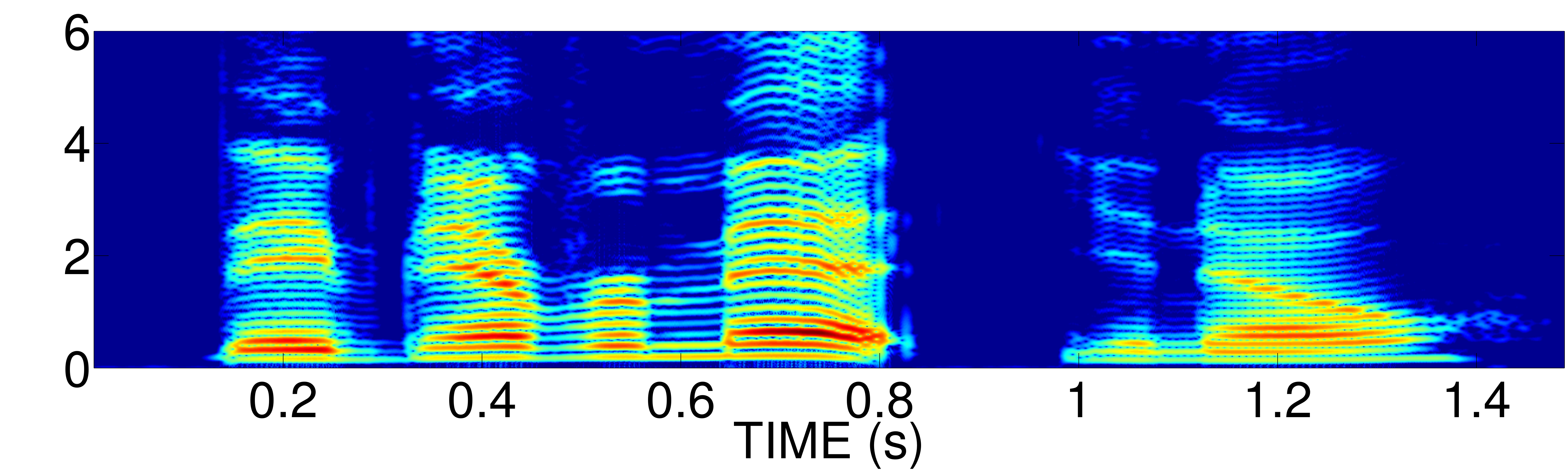}} \hspace{-0.03 in}{(c)}
   \end{tabular}
  \caption{[Color online] Spectrograms of the utterance ``they never met, you know''. (a) Original speech signal;  pitch-scaled signals for (b)  $\beta=1.2$ and (c) $\beta=0.8$.}
  \label{fig:PSM_result_spec}
 \end{figure}

\begin{figure}
  \centering
  \includegraphics[width= 2.35 in]{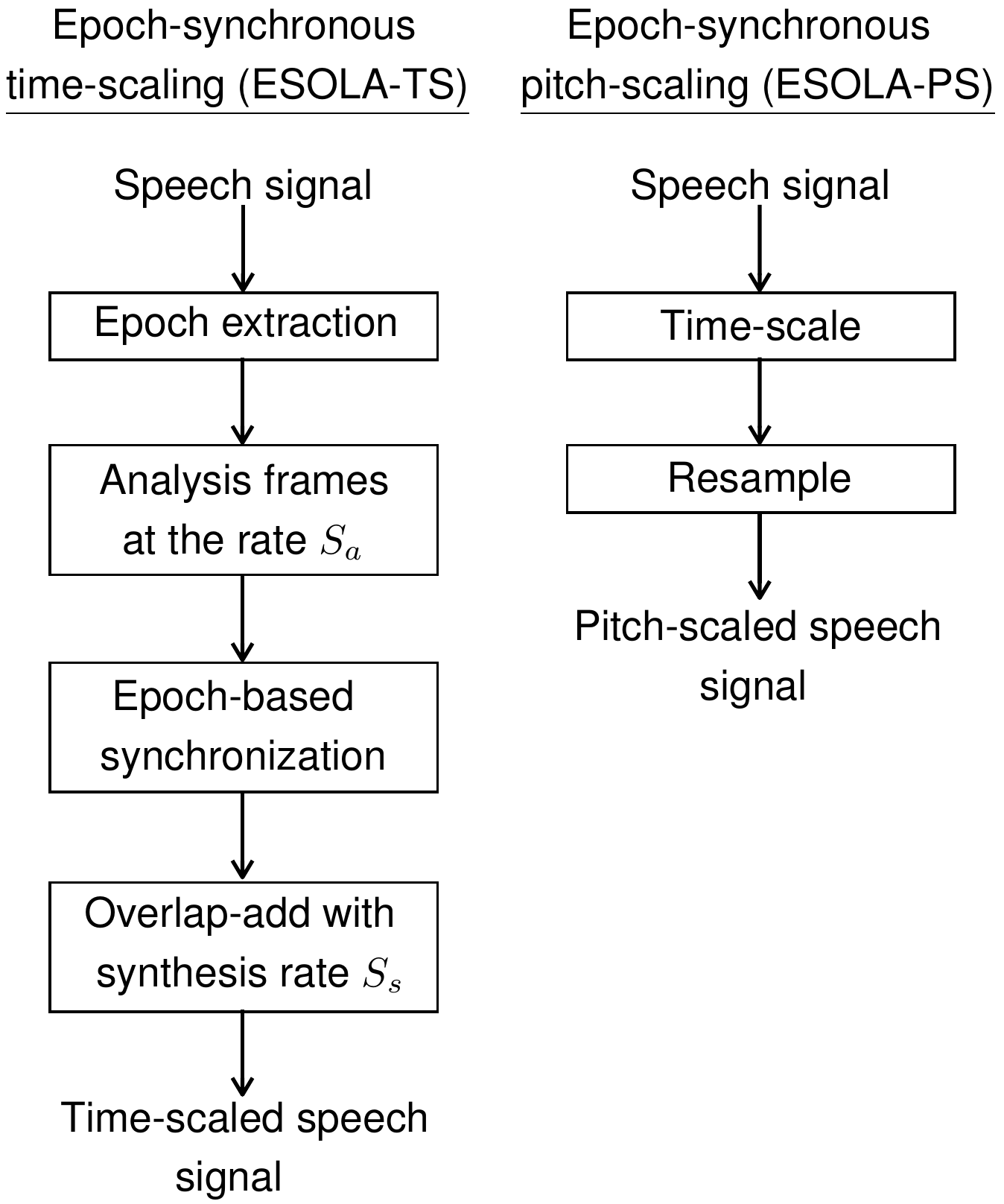} 
  \caption{Block diagram of ESOLA-TS/PS techniques.}
  \label{fig:flowchart}
 \end{figure}

Figs. \ref{fig:TSM_result_spec} and \ref{fig:PSM_result_spec} show the spectrograms of the speech signal corresponding to the utterance ``they never met, you know'' and its time- and pitch-scaled versions. It is observed that the spectral contents in the time-scaled spectrograms are preserved and do not contain any significant artifacts.  The proposed time- and pitch-scale modification techniques using ESOLA are summarized in the form of flowcharts in Fig. \ref{fig:flowchart}.

\section{Epoch Embedding}
\label{sec:epoch_embed}
The significant computation involved in time- and pitch-scaling methods is in the evaluation of measures providing synchronization between successive frames. Generally, normalized autocorrelation function, spectral autocorrelation function, short-time energy, etc. are utilized as measures for synchronization \cite{sola, solafs, 319366, Wong03fastsola-based, Laprie2}. These measures are to be computed for each analysis frame repeatedly for different time- and pitch-scale 
modification factors as they change with varying frame lengths and shifts. But epochs in speech signals are invariable to the changes in segmentation lengths and also to different time-scale modification factors. Hence, epochs can be extracted once from the speech signal and can be used repeatedly for different tasks. Thus the proposed method allows one to exploit this property to further reduce the computations involved. 

To exploit the advantage of the fact that epochs could be extracted once and be used repeatedly for different scale-factors, we propose the method of epoch embedding into the speech signal. Consider an array of all zeros, whose length is equal to the length of the signal. Now, the values at the sample indices corresponding to the epochs  are set to $1$. This array of binary decision about presence or absence of epochs is used to change the least significant bit (LSB) in the $8/16$-bit representation of the speech samples. If epoch is not present in the speech sample under consideration, then the LSB of that sample is set to $0$. If the speech sample indeed represents an epoch, then its LSB is set to $1$. 
Thus the epochs  are computed and saved for further use in time- and pitch-scale modifications to different factors. This strategy largely reduces the computational cost and execution time.

\section{Comparison of Different Time- and Pitch- Scaling Techniques}
\label{sec:compn}
In this section, we discuss the key differences between the proposed methodology and state-of-the-art time and pitch-scale modification techniques.  We broadly divide the existing techniques in the literature into two classes: (i) pitch-synchronous windowing techniques and (ii) pitch-blind windowing techniques.
   \begin{table*}[t]
  \centering
  \caption{Execution Time (in Seconds) of Different Time-Scale Modification Techniques for a Speech Signal of Duration $12$ Seconds}
  \label{tab:exectime}
  \begin{tabular}{|c|c|c|c|c|c|}
  \hline
  Time-scale factor  & TD-PSOLA \cite{Moulines} & LP-PSOLA \cite{MOULINES1995175} & WSOLA \cite{wsola} & SOLAFS \cite{solafs} & ESOLA  \\ \hline
 $0.5$ & 74.16 & 74.15 & 1.83 & 1.73 &  0.65 \\ \hline
 $0.75$ & 74.25 & 74.19 & 3.22 & 2.58 & 0.66 \\ \hline
 $1.25$ & 74.32 & 74.30 & 2.71 & 4.30 & 0.69 \\ \hline
 $1.5 $ & 74.38 & 74.35 & 2.37 & 5.13 &  0.70\\ \hline
 $2 $ &74.41 & 74.45 & 4.59 & 6.89 &  0.91\\ \hline
  \end{tabular}
 \end{table*}
\subsubsection{Pitch-Synchronous Windowing Techniques} 
The PSOLA and its variants mainly constitute the class (i) of techniques. As discussed in Section \ref{sec:pitch_sync_tech}, these class of techniques employ pitch synchronous windowing of speech signals, where each window is centered around pitch markers and typically covers two pitch periods \cite{Moulines}. Generally, tapered windows like Hamming or Hann windows are used for short-time segmentation of speech. Such windowed segments are replicated or deleted appropriately for time-scale modification, and are resampled for pitch-scale modification \cite{Moulines, MOULINES1995175}. 
 
The key philosophy behind the proposed ESOLA method is different from the PSOLA-based techniques in the sense that we employ pitch-blind windowing to segment the speech signal, where each segment grossly holds three to four pitch periods. The overlap between adjacent segments are manipulated in a controlled fashion for time-scale modification. The time-scaled speech is resampled appropriately for pitch-scale modification.  As segmentation in ESOLA method is simpler and the number of frame manipulations required is lesser than PSOLA leading to computationally efficient method that produces superior quality time and pitch-scaled speech. 
 
Also, pitch-synchronous methods are not able to produce exact time-scaled speech signals unlike the proposed ESOLA method, which delivers exact time-scale modifications due to fixed synthesis strategy.
 
 \subsubsection{Pitch-Blind Windowing Techniques} 
As detailed in Section  \ref{sec:pitch_blind_tech}, the SOLA and its variants adopt pitch-blind windowing of speech signals for short-time segmentation, and adjusts the overlap between  successive frames based on some synchronization measure for time-scale modification \cite{sola}. The  ESOLA method follows the same philosophy and yet the specific advantages delivered by the proposed method in terms of computational requirements and execution time make it superior to the other methods.

 The SOLA algorithm and a wide range of its variants use autocorrelation measures for frame synchronization \cite{sola, 379979, lawlor}, which has to be repeatedly computed for different time and pitch scaling factors adding up to the total computational cost and execution time. The variants of SOLA employing synchronization of frames
 using AMDF, mean-square differences, envelope matching, peak alignment, etc. \cite{319366, Wong03fastsola-based, 5745327, 1198877} are highly susceptible
 to noise. Epoch-based synchronization and epoch embedding proposed in ESOLA method contribute to reduction of overall computational requirements and since epochs are the high energy content in speech signals delivering high signal to noise ratio in regions around it \cite{enhance}, the resulting time and pitch-scaled speech signals are relatively robust to noise and holds superior perceptual quality. 
  
 To indicate the computational advantages rendered by ESOLA algorithm over the existing techniques, we tabulate the execution time required by different time-scale modification algorithms in Table~\ref{tab:exectime}. The reported execution times for the ESOLA method include the computation times involved in extracting  epoch locations also. In this study, we have used ZFF algorithm \cite{ZF} to estimate epoch locations. The The codes were run in MATLAB R2015 on a Macintosh computer equipped with 2.7 GHz Intel Core i5 processor, 8 GB 1333 MHz DDR3 RAM.  The ESOLA algorithm is the fastest among the algorithms under consideration, bringing out its advantage in applications to real time systems.

  \begin{table}
  \centering
  \caption{Attributes Considered to Rate a Time/Pitch-Scaled Speech}
  \label{tab:quality}
  \begin{tabular}{|l|l|}
  \hline
  No: & Attributes \\ \hline
  1 & Intended changes, i.e., whether the duration/speed or pitch \\ 
    & of speech files has indeed changed or not \\ \hline
  2 & Pitch consistency in time-scale modification \\ \hline
  3 & Duration consistency in pitch-scale modification \\ \hline
  4 & Perceptual quality and intelligibility \\ \hline
  5 & Distortions or artifacts \\ \hline
  \end{tabular}
 \end{table}
 \begin{table}
  \centering
  \caption{Ratings Used for Assessing the Quality of Time/Pitch-Scaled Speech \cite{KSRao}}
  \label{tab:rating}
  \begin{tabular}{|l|l|l|}
  \hline
  Rating & Speech quality & Distortions \\ \hline
  1 & Unsatisfactory & Very annoying and objectionable \\ \hline
  2 & Poor & Annoying, but not objectionable \\ \hline
  3 & Fair & Perceptible and slightly annoying \\ \hline
  4 & Good & Just perceptible, but not annoying \\ \hline
  5 & Excellent & Imperceptible \\ \hline
  \end{tabular}
 \end{table}
 \section{Perceptual Evaluation}
 \label{sec:eval}
 To evaluate the performance of the proposed ESOLA method in comparison with other state-of-the-art techniques, we conducted detailed perceptual evaluation tests. Three English speech utterances of $3$-$4$ seconds duration, two spoken by a male speaker and one by a female speaker, are chosen from the CMU Arctic database \cite{cmu}. The speech signals are down sampled to $16$ kHz and are segmented into frames of $20$ ms duration with synthesis rate ($S_s$) of $10$ ms for time and pitch-scale modification. The time and pitch scaling are performed for five different modification factors as mentioned in Table.~\ref{tab:time} and Table.~\ref{tab:pitch}, respectively. All three sentences are time and pitch-scaled for the chosen modification factors making a total of three sets of speech files for perceptual evaluation. Twenty five listeners with a basic understanding of speech signal processing, notions of pitch, duration, playback rate, etc. were chosen for the evaluation test. Each listener was asked to listen carefully to one set of speech files and the three listening sets were randomly distributed among the $25$ listeners, in order to remove any bias in evaluation to a particular speaker or utterance. 
 \begin{table}[t]
  \centering
  \caption{MOS for Different Time-Scale Modification Algorithms}
  \label{tab:time}
  \begin{tabular}{|c|c|c|c|c|c|}
  \hline
  Time-scale & TD- & LP- & WSOLA & SOLAFS & ESOLA\\
  factor & PSOLA & PSOLA & & &  \\ \hline
    $0.5$ & 1.58 & 2.11 & 4.15 & 4.15 & \textbf{4.21} \\ \hline
  $0.75$  & 1.58 & 2.16 & 3.84 & 4.26 & \textbf{4.53} \\ \hline
  $1.25$ & 1.80 & 2.35 & 3.45 & 4.30 & \textbf{4.80} \\ \hline
  $1.5$ & 1.55 & 2.20 & 3.25 & 4.10 & \textbf{4.65} \\ \hline
  $2$ & 1.60 & 2.20 & 3.10 & 3.85 & \textbf{4.15} \\ \hline
  \end{tabular}
 \end{table}
 \begin{table}[t]
  \centering
  \caption{MOS for Different Pitch-Scale Modification Algorithms}
  \label{tab:pitch}
  \begin{tabular}{|c|c|c|c|c|c|}
  \hline
  Pitch-scale & TD- & LP- & WSOLA & SOLAFS & ESOLA \\
  factor & PSOLA & PSOLA & & & \\ \hline
$0.5$ & 1.40 & 1.50 & 2.40 & 3.95 & \textbf{4.25} \\ \hline
  $0.75$ &1.60 & 1.45 & 2.60 & 4.25 & \textbf{4.35} \\ \hline
  $1.25$ & 1.75 & 1.50 & 3.30 & 4.20 & \textbf{4.60} \\ \hline
  $1.5$ & 2.20 & 1.85 & 3.00 & 4.25 &\textbf{4.40} \\ \hline
  $2$ &2.40 & 1.60 & 2.35 & 4.30 & \textbf{4.40} \\ \hline
  \end{tabular}
 \end{table}
The listeners were asked to rate each speech file on a scale of $1$ to $5$ based on the attributes given in Table.~\ref{tab:quality}. Each point in the rating represents the speech quality and level of distortion as given in Table.~\ref{tab:rating} \cite{KSRao}. Each listener took approximately $45$ minutes to complete the task of evaluation.
 \begin{figure}[t]
  \centering
   \includegraphics[width=\columnwidth]{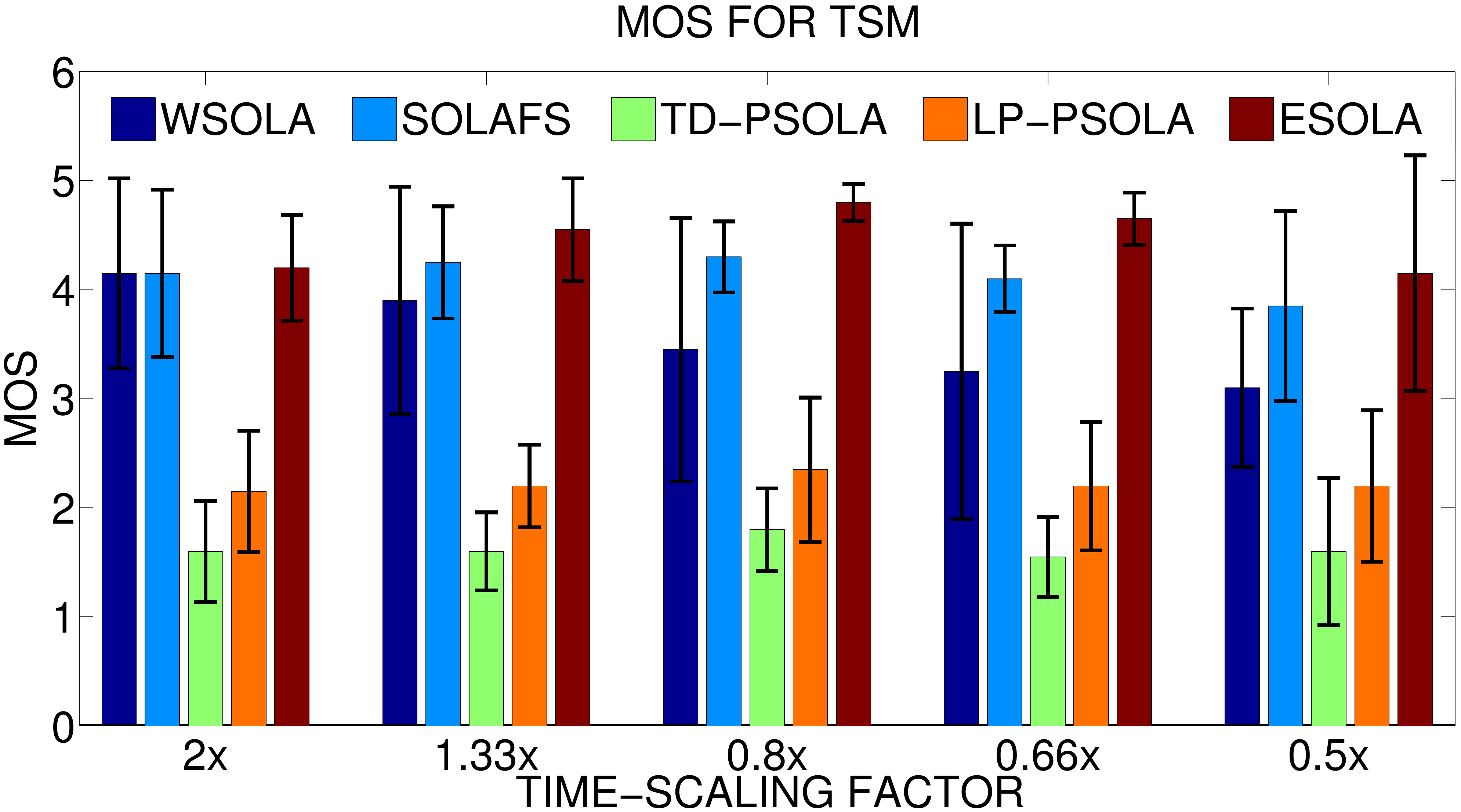}
  \caption{[Color online] MOS of various time-scaling techniques for five different scaling factors. Variance of MOS for a TSM technique and a particular scaling factor is also plotted as a vertical line on top of the bar graph.}
  \label{fig:TSM_MOS}
 \end{figure}
  \begin{figure}[t]
  \centering
  \includegraphics[width=\columnwidth]{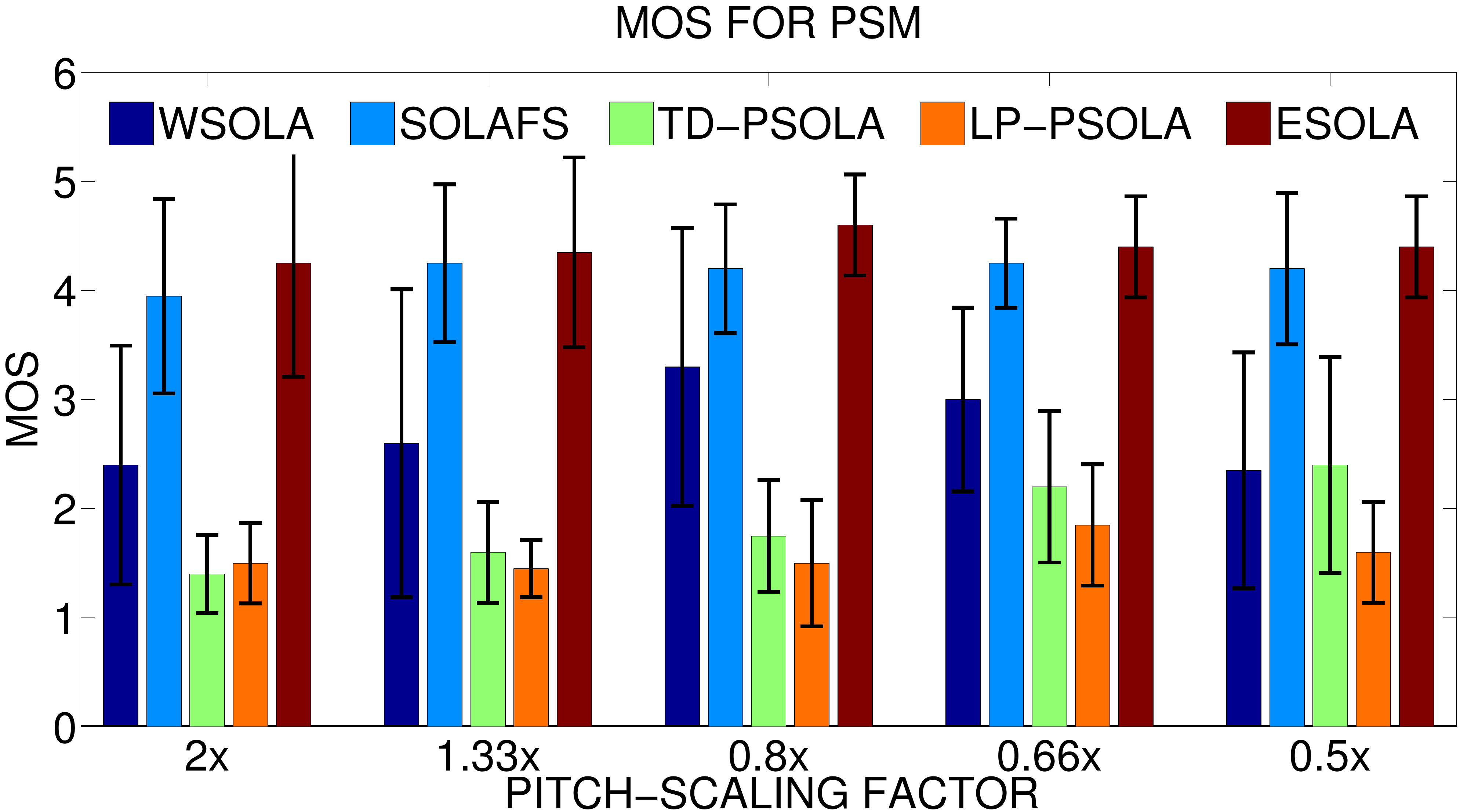}
  \caption{[Color online] MOS of various pitch-scaling techniques for five different scaling factors. Variance of MOS for a PSM technique and a particular scaling factor is also plotted as a vertical line on top of the bar graph.}
  \label{fig:PSM_MOS}
 \end{figure}
For perceptual evaluation of performances, we have included four prominent time and pitch scaling methods reported in literature, namely, TD-PSOLA \cite{Moulines},  LP-PSOLA \cite{MOULINES1995175, KSRao}, WSOLA \cite{wsola}, and SOLAFS \cite{solafs}. The performances are computed as mean opinion scores (MOS) from $25$ listeners over all three listening sets of speech signals. The time and pitch scaling performance of different algorithms are given in Table.~\ref{tab:time} and Table.~\ref{tab:pitch}, respectively. Figs. \ref{fig:TSM_MOS} and \ref{fig:PSM_MOS} show the performance results as bar graphs along with the variances in MOS of $25$ listeners. The ESOLA algorithm consistently delivers better MOS values than the rest indicating the better quality of time/pitch-scaled speech of the proposed technique. Next, we list the observations based on the results of perceptual evaluation. 
\begin{itemize}
\item The ESOLA method significantly outperforms the PSOLA-based techniques in both time and pitch scaling. This could be attributed to the simpler and efficient frame manipulations in the ESOLA algorithm. The performance of LP-PSOLA is poor compared with TD-PSOLA in pitch scaling because of the filtering in LP residue domain.
\item The ESOLA  method achieves exact time scaling of speech signals. Whereas the WSOLA is not capable of producing speech exactly for a specified time-scale factor and it loses duration consistency in pitch-scale modification, owing to the degraded time and pitch scaling performance. 
\item The SOLAFS and ESOLA algorithm provide comparable performances with ESOLA having an edge over the SOLAFS method. This could be attributed to the fact that the ESOLA method performs frame alignment based on epoch information, which is more accurate than frame alignment based on cross correlation analysis as done in SOLAFS. 
\item One of the positives of ESOLA method is its computational efficiency over other methods as discussed in Section~\ref{sec:compn} (Table \ref{tab:exectime}). The proposed method reserves the least execution time among other prominent methods.
\end{itemize}

    
 \section{Conclusions}
 \label{sec:concl}
 In this paper, we proposed a computationally efficient, real-time implementable, and superior quality time and pitch-scale modification algorithms.
 The proposed technique (ESOLA) employs short-time segmentation of speech signals using pitch-blind windowing and manipulates the overlap between 
 successive frames for time-scale modification. Appropriate resampling of time-scaled speech is performed for pitch-scale modification. The key features
 of the ESOLA algorithm are the utilization of epochs for synchronization of successive frames to remove pitch inconsistencies, deletion or insertion 
 of samples in synthesis frames to ensure fixed synthesis and the technique of epoch embedding to significantly reduce the computational cost. Subjective
 experiments conducted to study the performance of different time and pitch scaling algorithms revealed the superiority of the proposed technique.
 The ESOLA algorithm significantly outperforms PSOLA based techniques due to its simpler and efficient frame manipulations, and the SOLAFS due to 
 the accurate frame alignment based on epochs. Also, the ESOLA algorithm is computationally efficient and requires the least execution time for its 
 implementation among the other prominent time and pitch scaling techniques.
  
\section{Acknowledgements}
 The authors would like to thank Jitendra Dhiman (C++), Pavan Kulkarni (Android application), Aishwarya Selvaraj (Praat plugin), and Vinu Sankar (Python) for implementing the ESOLA algorithm with nice GUIs on various platforms.

 \bibliographystyle{IEEEtran}
\bibliography{IEEEabrv,ets-epsRefs}

\end{document}